\documentclass[smallextended,12pt]{svjour3}

\usepackage[export]{adjustbox}
\usepackage[utf8]{inputenc}
\usepackage{mathpazo}
\usepackage{amsmath}
\usepackage[round]{natbib}
\usepackage[margin=1in]{geometry}
\usepackage{amssymb}
\usepackage{standalone}
\usepackage{graphicx}
\usepackage{overpic}
\usepackage{epstopdf}
\usepackage[hyperfootnotes=false,colorlinks=true,allcolors=black,urlcolor=blue,anchorcolor=blue]{hyperref}
\usepackage[capitalise]{cleveref}
\usepackage{isomath}
\usepackage[export]{adjustbox}
\usepackage{caption}
\usepackage[normalem]{ulem}
\usepackage[rgb,dvipsnames]{xcolor}
\usepackage{tikz}
\usepackage{bm}
\usepackage[caption=false]{subfig}
\usepackage{booktabs}
\usepackage[math]{cellspace}
\usepackage{multirow}

\AtBeginDocument{
    \heavyrulewidth=.08em
    \lightrulewidth=.05em
    \cmidrulewidth=.03em
    \belowrulesep=.65ex
    \belowbottomsep=0pt
    \aboverulesep=.4ex
    \abovetopsep=0pt
    \cmidrulesep=\doublerulesep
    \cmidrulekern=.5em
    \defaultaddspace=.5em
}

\definecolor{MyBlue}{rgb}  {0.1,0.1,0.9}
\definecolor{MyRed}{rgb}   {0.9,0.1,0.1}
\definecolor{MyGreen}{rgb} {0.05,0.4,0.05}
\definecolor{burntorange}{rgb}{0.8, 0.33, 0.0}
\definecolor{NeilMagenta}{rgb}{0.8, 0.1, 0.8}

\newcommand{\ak}[1]{#1}

\DeclareSymbolFont{otone}{OT1}{pplx}{m}{n}
\DeclareMathSymbol{\Delta}{\mathalpha}{otone}{1}

\crefformat{footnote}{#2\footnotemark[#1]#3}

\newcommand{\pd}[2]{\frac{\partial #1}{\partial #2}}

\newcommand{\norm}[1]{\left\lVert#1\right\rVert}
\newcommand{\vnabla}{\boldsymbol{\nabla}}
\newcommand{\lap}{\nabla^2}

\newcommand{\beq}{\begin{equation}\begin{aligned}}
\newcommand{\eeq}{\end{aligned}\end{equation}}
\newcommand \beqno{\begin{eqnarray*}}
\newcommand \eeqno{\end{eqnarray*}}
\newcommand \bit{\begin{itemize}}
\newcommand \eit{\end{itemize}}

\begin{document}
\title{Turing instabilities are not enough to ensure pattern formation}
\titlerunning{Turing instabilities are not enough}
\author{Andrew L.\ Krause\footnote{Corresponding author \email{andrew.krause@durham.ac.uk}}\and Eamonn A.\ Gaffney \and Thomas Jun Jewell \and V{\'a}clav Klika \and  Benjamin J.\ Walker}
\authorrunning{Krause et al}

\institute{
    A. L. Krause
    \at Department of Mathematical Sciences, Durham University, Upper Mountjoy Campus, Stockton Road, Durham DH1 3LE, United Kingdom \\
    E. A. Gaffney, T. J. Jewell
    \at Wolfson Centre for Mathematical Biology, Mathematical Institute, University of Oxford, 
    Oxford, OX2 6GG, United Kingdom
    \\
    V. Klika
    \at Department of Mathematics, FNSPE, Czech Technical University in Prague, Trojanova 13, 120 00 Praha, Czech Republic
    \\
    B. J. Walker
    \at Department of Mathematical Sciences, University of Bath, Bath BA2 7AY, United Kingdom; Department of Mathematics, University College London, London WC1E 6BT, United Kingdom
    }

\date{Received: date / Accepted: date}

\maketitle

\begin{abstract}
Symmetry-breaking instabilities play an important role in understanding the mechanisms underlying the diversity of patterns observed in nature, such as in Turing's reaction--diffusion theory, which connects cellular signalling and transport with the development of growth and form. Extensive  literature focuses on the linear stability analysis of homogeneous equilibria in these systems, culminating in a set of conditions for transport-driven instabilities that are commonly presumed to initiate self-organisation. We demonstrate that a selection of simple, canonical transport models with only mild multistable non-linearities can satisfy the Turing instability conditions while also robustly exhibiting only transient patterns. Hence, a Turing-like instability is insufficient for the existence of a patterned state. \ak{While it is known that linear theory can fail to predict the formation of patterns, we demonstrate that such failures can appear robustly in systems with multiple stable homogeneous equilibria.} Given that biological systems \ak{such as} gene regulatory networks and spatially distributed ecosystems often exhibit a high degree of multistability and nonlinearity, this raises important questions of how to analyse prospective mechanisms for self-organisation.
\end{abstract}
 

\section{Introduction}
Nature exhibits  diverse  structures in the organisation of life across spatial and temporal scales. Elaborate animal coat patterns \citep{koch1994biological}, emergent territory boundaries between predators \citep{potts2016memory}, and complex spatiotemporal arrangements of slime moulds \citep{hofer1995cellular} are a few of the patterns researchers have sought to understand. A key mechanism underlying such patterns are symmetry-breaking (Turing) instabilities of spatially uniform equilibria, as explored in Turing's influential \emph{Chemical basis of morphogenesis} \citep{turing1952chemical}.

Typical analysis of these phenomena is often based on linear stability theory, which attempts to ascertain the growth or decay of perturbations to homogeneous equilibria. Due to the nature of the resulting linear equations, such analysis can often be carried out very easily. In addition to its simplicity, a chief advantage to this approach is its generality, as it makes minimal assumptions about the precise form of the underlying system. In turn, this provides  reasonably broad statements about the kinds of systems that can exhibit such instabilities, as illustrated by  the fact that Turing self-organisation in a two-species reaction diffusion system requires a short-range (self)-activator, and a long-range (self)-inhibitor \citep{meinhardt2000pattern}. The simplicity of linear stability analysis means that, even for many-species systems \citep{marcon2016high}, one can typically classify parameters of the linearised system into those that exhibit pattern-forming instabilities (the so-called `Turing space'), and those which cannot \citep{murray1982parameter, murray2003mathematical}. There is a large body of work aimed at understanding features of these Turing spaces in various contexts \citep{klika2017history, marcon2016high, gaffney2023spatial}, but always using some form of linear stability theory, which is a dominant feature of the pattern formation literature. Hence a large number of studies have focused on linear systems exclusively to make general claims about proposed mechanisms \citep{satnoianu2000turing,krause2020one, haas2021turing} or to design pattern-forming systems with certain properties \citep{vittadello2021turing,woolley2021bespoke}.

However, when linear analysis identifies a pattern-forming instability, the output is always a local result, in that transient symmetry-breaking patterns are expected to form from perturbations of the homogeneous steady state. Beyond the formation of an initial pattern, linear stability provides no guarantee of a long-time (i.e.~stable) patterned state. \ak{Notably, the existence of stable patterns can be guaranteed in the case of supercritical Turing bifurcations, but only near the boundary of the Turing space \citep{vastano1988turing}}. However, the emergence of large-scale, persistent self-organisation is invariably presumed from the linear analysis (including by the authors), often based on intuition and experience with simple examples of minimal complexity \citep{murray2003mathematical, krause2021modern}.   

While this intuition has been seen to be correct for many textbook systems, extensive recent examples highlight that linear stability theory cannot always capture the fundamental dynamics of pattern-forming systems, such as instabilities due to subcritical bifurcations \citep{champneys2021bistability, villar2023degenerate}. Unlike in the supercritical case, subcritical bifurcations do not typically admit small-amplitude stable patterned states, even in the weakly nonlinear regime except very near to the codimension-2 point where the criticality of the bifurcation changes \citep{brena2014subcritical}. \ak{Such subcritical bifurcations can lead to pattern formation outside of Turing space, as implicated in ecological work on resilience due to patterning \citep{van2004spatial, bastiaansen2020effect}, among other areas. Subcritical bifurcations can also lead to spatiotemporal oscillations and chaos \citep{painter2011spatio}. Other secondary bifurcations can eliminate any stable patterned branches, so that systems with multiple spatial homogeneous equilibria may form only transient patterned states; see Figures 8 and 11 in \citet{al2020bloom} for an example. Non-normality (in the sense of normal matrices/operators) can also lead to different predictions from linear theory, as described by \cite{klika2017significance}.} Thus, classical linear stability conditions are \ak{neither necessary nor sufficient} for self-organisation. 

Here, we demonstrate that \ak{this insufficiency of the} classical Turing conditions \ak{can occur generically in a range of systems}. In particular, we exemplify that the presence of multistability can robustly spoil typical predictions of patterning by driving a system to a stable homogeneous equilibrium after the emergence of transient patterns via a Turing instability. \ak{Multistability has become an increasingly prominent topic in gene regulatory networks \citep{laurent1999multistability, siegal2009capacity, feng2016core, bocci2023theoretical}, ecology \citep{suzuki2021energy}, and evolutionary biology \citep{arnoldt2012frequency}, with growing evidence that multistable dynamics are ubiquitous in biological systems. Here, we show that even bistability of reaction kinetics can alter the prospect for pattern formation in a robust way, suggesting a need for better tools to analyze more realistic models of pattern formation in biological systems.} 

\ak{The rest of the paper is organized as follows.} In \cref{Models}, we present and \ak{perform a linear stability analysis of specific} models from four distinct classes of pattern-forming systems. In \cref{Results}, we perform thousands of numerical experiments with random parameters and demonstrate that these models void our long-established intuition for pattern formation relying on linear stability theory, raising important issues regarding the connection between textbook analyses and realistic biological systems, which we discuss in \cref{Discussion}.

\section{Models \& dispersion relations}\label{Models}

We consider four models on the spatial domains $\Omega = [0,L]$ or $\Omega = [0,L]\times [0,L]$, with periodic boundary conditions. Parameters are assumed to take positive nonzero values, with exemplars given in \cref{table_params} for each model, which we will refer to as the \emph{base parameters}. 

For each model, we perform a linear stability analysis around one of the spatially homogeneous equilibria and record the growth rate of spatial perturbations corresponding to the eigenvalues $\rho_k$ of the negative Laplacian given by
\beq
\lap w_k(\bm{x}) = -\rho_k w_k(\bm{x})
\eeq
with periodic boundary conditions, ordered via $0=\rho_0< \rho_1 \leq \rho_2 \leq \dots$, with $w_k$ the corresponding eigenfunctions (these are just sinusoidal functions for these domains and boundary conditions).
We then write the maximal growth rate of linear perturbations corresponding to eigenfunction $w_k$ as $\lambda_k$, so that $\Re(\lambda_0)<0$  and $\Re(\lambda_k)>0$ for some $k>0$ is our criterion for a Turing instability. Analysing the first three models is standard \citep{murray2003mathematical, krause2021modern}, whereas the linear stability theory for the nonlocal advection model is given by \cite{jewell2023patterning}. \ak{In each case, we will focus on the linear stability of one equilibrium, but each model will also admit one other stable equilibrium.}

\begin{table}
\centering
    \begin{tabular}{| l | c | c | c | c | c | c | c | }
    \hline
    Model   &$L$ &$D$ & $a$ & $b$ &$c$ & $d$ & $e$  \\ \hline
    Reaction--diffusion  &100 &$30$ & $1.75$ & $18$ &$2$ & $5$ & $0.02$  \\ \hline
    Keller--Segel   &$80$ &$1$ & $1$ & $1$ &$3$ & $0.8$ & -  \\ \hline
    Biharmonic  &$100$ & $1.45$  & $5$ & $0.9$ &$1$ & - & -  \\ \hline
    Nonlocal Advection  & $30$ &$1$ & $1$ & $0.45$ &$0.5$ & $20$ & -\\ \hline

    \end{tabular}
    \caption{\ak{Base} model parameters \ak{for the four different models}.}\label{table_params}
\end{table}

\subsection{Reaction--diffusion system}
We first consider a two-component reaction--diffusion system of the form
\beq\label{RD_model}
&\pd{u}{t} = \lap u + u - v - e u^3,\\
&\pd{v}{t} = D \lap v + a v(v + c)(v - d) +  b u - e v^3,
\eeq
which has a homogeneous equilibrium at $(u_0,v_0)=(0,0)$.
For the parameters in \cref{table_params}, this equilibrium is 
stable in the absence of diffusion. There are four   further real equilibria,  { only one of which is stable in the absence of diffusion.}  Linearising \cref{RD_model} around $(u_0,v_0)=(0,0)$ gives perturbation growth rates
\beq
        \lambda_k = \frac{1-acd-\rho_k(1+D)+\sqrt{(1-acd-\rho_k(1+D)))^2-4(\rho_k^2 D-\rho_k(D-acd)-acd+b))}}{2}.
\eeq
\ak{For the base parameters, the equilibrium $(u_0,v_0)=(0,0)$ is Turing unstable (see the plot of the dispersion relation in \cref{fig_1D_dispersion_sims}(a)).}

\subsection{Keller--Segel with Allee demographics}
We next consider a Keller--Segel \citep{keller1970initiation, horstmann20031970} model of chemotaxis:
\beq\label{KS_model}
\pd{u}{t} &=  \lap u - c\vnabla \cdot(u\vnabla v) +u(b - u)(u - d),\\
\pd{v}{t} &= D \lap v + u-av.
\eeq
The system admits three spatially homogeneous equilibria, $v_0=u_0/a$ with $u_0=0, d, b$. This is bistable in the absence of transport if $b > d > 0$, with stable equilibria $u_0=0$ and $u_0=b$. Linearising \cref{KS_model} around $(u_0,v_0)=(b,b/a)$ gives perturbation growth rates
\beq
        \lambda_k = \frac{T_\textrm{KS}+\sqrt{T_\textrm{KS}^2-4Q_{KS}}}{2},
\eeq
where $ T_\textrm{KS} =-b(b-d)-a-\rho_k(1+D)$, and $Q_{\textrm{KS}} = \rho_k^2 D-\rho_k(cb-a-Db(b-d))+b(b-d)a$. \ak{For the base parameters, the equilibrium $(u_0,v_0)=(b,b/a)$ is Turing unstable  (see the plot of the dispersion relation in \cref{fig_1D_dispersion_sims}(b)).}

\subsection{Biharmonic instability}
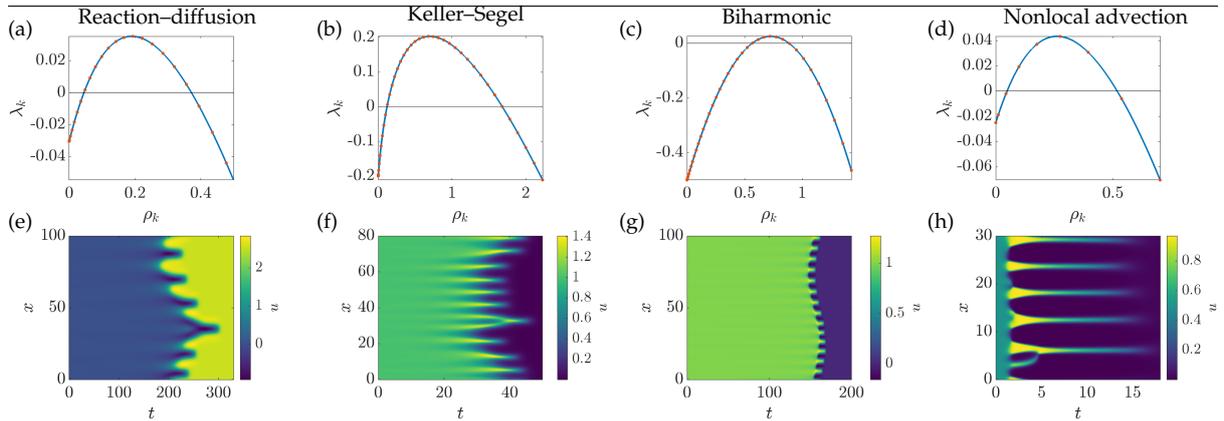
\begin{figure*}
    \centering
    \begin{overpic}[permil,width=\textwidth]{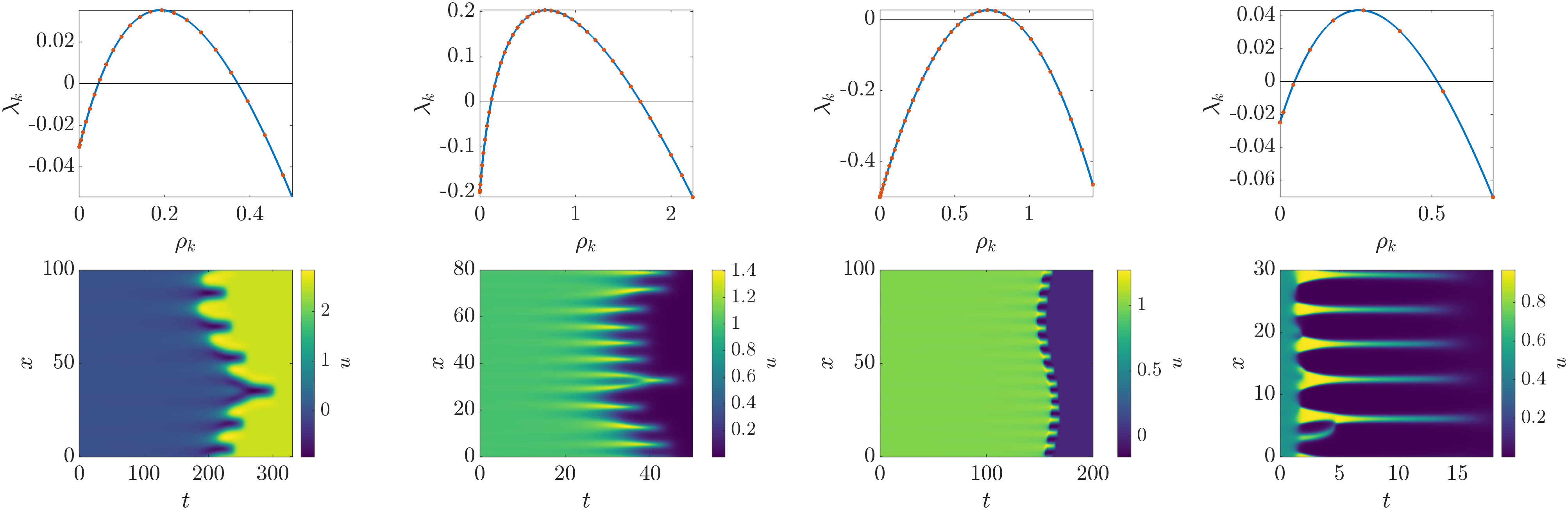}
        \put(0,320){(a)}
        \put(255,320){(b)}
        \put(505,320){(c)}
        \put(760,320){(d)}
        \put(0,160){(e)}
        \put(255,160){(f)}
        \put(505,160){(g)}
        \put(760,160){(h)}
        \put(58,330){{\footnotesize Reaction--diffusion}}
        \put(330,332){{\footnotesize Keller--Segel}}
        \put(590,330){{\footnotesize Biharmonic}}
        \put(823,330){{\footnotesize Nonlocal advection}}
    \end{overpic}
    \caption{(a)-(d) Plots of $\lambda_k$ against the continuous spatial eigenvalue $\rho_k$ in 1D, with orange dots corresponding to discrete values $\rho_k$ from the finite domains \ak{of size} $L$ \ak{given in \cref{table_params}, with the equilibrium being perturbed given in the text}. (e)-(h) Kymographs of $u$ over time in each model following perturbations from their Turing-unstable equilibria. Columns correspond to the models of \cref{Models}.}
    \label{fig_1D_dispersion_sims}
\end{figure*}

Next, we consider a fourth-order model of self-organisation:
\beq\label{Biharmonic_model}
\pd{u}{t} = - D\lap u -  \nabla^4 u + au(c - u)(u - b).
\eeq
The spatially homogeneous equilibria are $u_0=0,b,c$, which exhibit bistability \ak{of $u_0=0$ and $u_0=c$} in the absence of transport for $c > b > 0$. Linearising \cref{Biharmonic_model} around $u_0=c$ gives perturbation growth rates
\beq
        \lambda_k = D\rho_k-\rho_k^2+ac(b-c).
\eeq
\ak{For the base parameters, the equilibrium  $u_0 = c$ is Turing unstable  (see the plot of the dispersion relation in \cref{fig_1D_dispersion_sims}(c)).}

\subsection{Nonlocal advection}
Finally, we consider an integro-differential model of cell aggregation \citep{painter2015nonlocal, jewell2023patterning}:
\beq\label{Nonlocal_model}
\pd{u}{t} &= D \lap u + au(c - u)(u - b) \\&- d\vnabla \cdot \left (u(1-u)\int_\Omega \frac{\bm{s}}{\norm{\bm{s}}}\frac{\mathrm{e}^{-\norm{\bm{s}}}}{2\pi}u(\bm{x}+\bm{s})d\bm{s}^N \right ),
\eeq
where $d\bm{s}^N$ is the volume element for $N=1$ or $N=2$ spatial dimensions. We require $c > b > 0$ for stability of the spatially homogeneous equilibria $u_0=0,c$ in the absence of transport, while $u_0=b$ is unstable. Linearising \cref{Nonlocal_model} around the $u_0=c$ equilibrium gives perturbation growth rates
\beq
        \lambda_k = -ac(c-b)-D\rho_k + \frac{c(1-c) d \rho_k}{\pi^{2-N}(1 + \rho_k)^{\frac{N+1}{2}}}.
\eeq
\ak{In 1D and 2D with the base parameters, the equilibrium  $u_0 = c$ is Turing unstable  (see the plot of the 1D dispersion relation in \cref{fig_1D_dispersion_sims}(d)).}

\section{Results}\label{Results}

\begin{figure*}
    \centering
    \begin{overpic}[permil,width=0.7\textwidth]{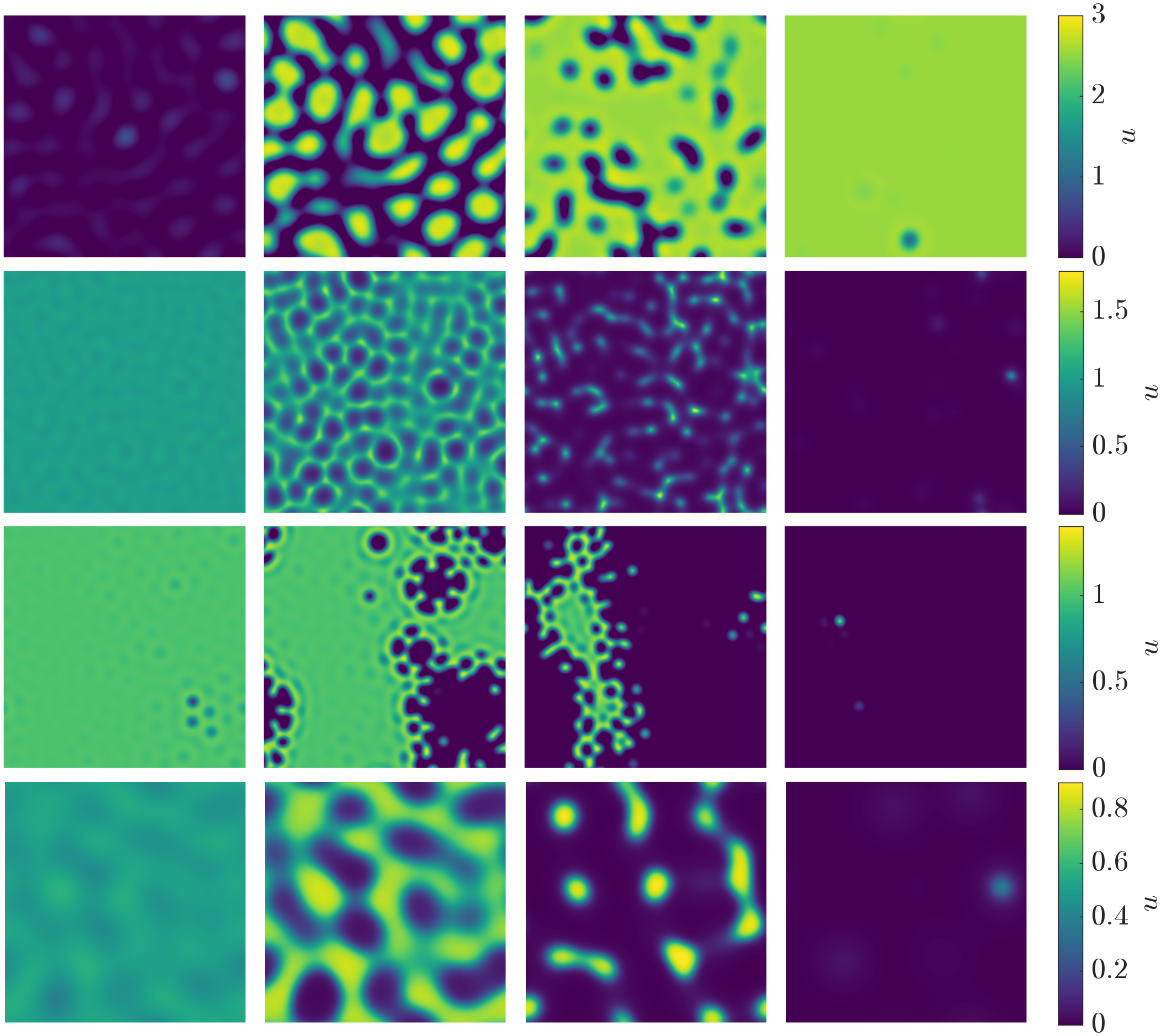}
    \put(-30,865){(a)}
    \put(-30,645){(b)}
    \put(-30,425){(c)}
    \put(-30,205){(d)}
    \put(400,-12){Time $\longrightarrow$}
    \put(-212,769){{\footnotesize Reaction--diffusion}}
    \put(-140,549){{\footnotesize Keller--Segel}}
    \put(-130,329){{\footnotesize Biharmonic}}
    \put(-217,109){{\footnotesize Nonlocal advection}}
    \end{overpic}
    \caption{Snapshots of transient 2D dynamics with Turing instabilities. (a)-(d) The evolution of $u$ for the models in \cref{Models} from initial perturbations. Simulations continue to decay towards homogeneous equilibria for times beyond those shown (see the Videos folder at \citet{Krause_Github}). The dynamics of (a)-(c) can be explored interactively using VisualPDE \citep{walker2023visualpde} at \url{https://visualpde.com/mathematical-biology/Turing-conditions-are-not-enough}.}
    \label{fig_2D_sims}
\end{figure*}
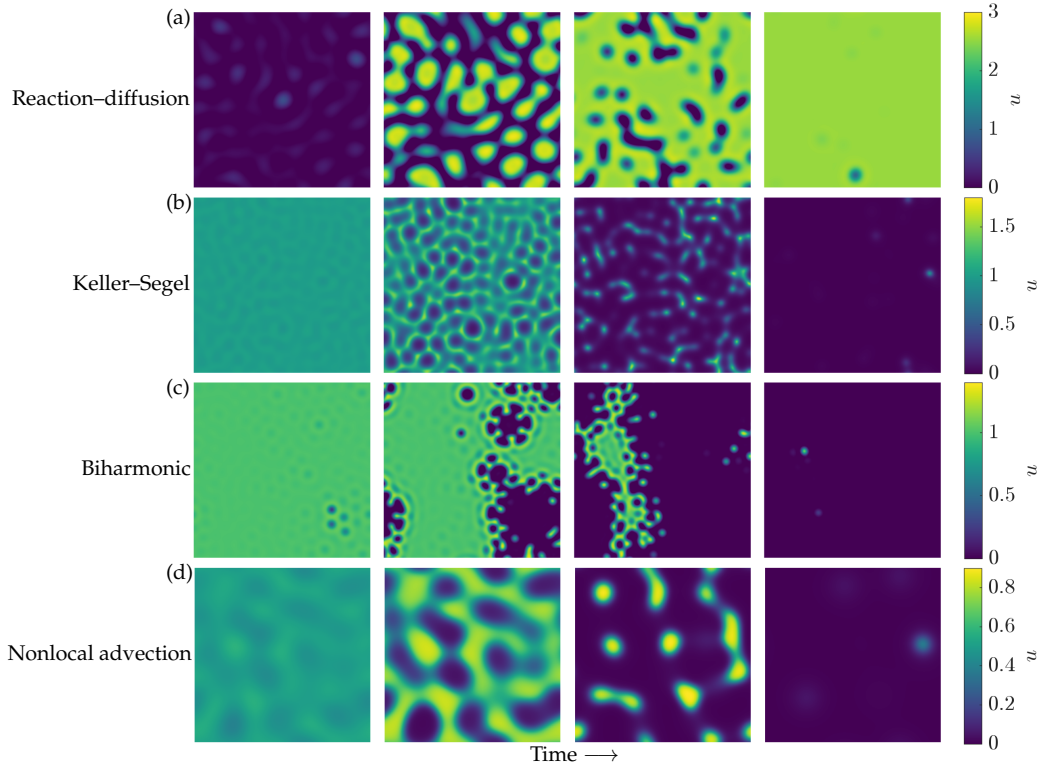

Each of these systems admits a Turing instability for the parameters given by \cref{table_params} for one of their equilibria, illustrated in the dispersion plots of \cref{fig_1D_dispersion_sims}(a)-(d). Hence, following commonplace reasoning, one might presume that a pattern \ak{(a stationary or spatiotemporal solution bounded away from homogeneous solutions)} will form from perturbations of these equilibria. However, numerical simulations of these models in 1D in \cref{fig_1D_dispersion_sims}(e)-(h) and in 2D in \cref{fig_2D_sims} show transient pattern formation that then decays to a different homogeneous equilibrium, all of which are linearly stable.  \ak{Briefly, the three local models are solved using finite differences, and the nonlocal model using a pseudospectral method combined with finite differences. In all cases, implicit timestepping algorithms are used, with initial data given by normally distributed perturbations of the Turing unstable equilibrium (of standard deviation $10^{-2}$), as detailed in the repository \cite{Krause_Github}.}

\ak{Importantly,} this decay to homogeneity occurs robustly \ak{across variation in all parameters}. To demonstrate this, we vary parameters and initial conditions as follows. For each model, we multiply every parameter given in \cref{table_params}, including the domain length $L$, by a uniformly random number from the interval $[0.95,1.05]$ using Latin Hypercube Sampling \citep{wyss1998users}. We then simulate the system for $t=10^4$ time units from a different random initial perturbation, recording both if there is a Turing instability (by analysing the dispersion relation) and if the system is approaching a homogeneous state (assessed by checking if  $\max_{\bm{x}}(u(10^4,\bm{x}))-\min_{\bm{x}}(u(10^4,\bm{x}))>10^{-5}$). From this, we can determine the proportion of simulations that exhibited a Turing instability but only patterned transiently from a small random perturbation of the homogeneous equilibrium. We perform $10^4$ simulations for all 1D models, and $10^3$ simulations for the 2D models. We omit the 2D nonlocal advection system from this analysis due to its numerical complexity.

We give results of this analysis in \cref{table_systematic_runs}. For all but the reaction--diffusion model, an overwhelming majority of cases converged to homogeneous equilibria after transiently patterning via a Turing instability. All of these systems remained within the Turing space of their corresponding equilibria. The reaction-diffusion model also exhibited robust convergence to a homogeneous equilibrium, though as the base parameters of the reaction--diffusion model lie near the boundary of the Turing space, not all simulations were Turing-unstable. A small proportion of the reaction--diffusion and Biharmonic systems were attracted to patterned equilibria, some of which were domain-filling while others appeared spatially localised \citep{champneys2021bistability}. Both the Keller--Segel and 1D nonlocal advection models only exhibited convergence to a homogeneous equilibrium after transient patterning.

\begin{table*}
\centering
    \begin{tabular}{| l | c | c | c | c |}
    \hline
    \multicolumn{2}{|l|}{Model} & Final state unpatterned & Turing unstable  & Turing conditions insufficient  \\ \hline
    \multirow{2}{*}{Reaction--diffusion} & 1D & $93.01$\% & $49.76$\%  & $86.17$\% \\ \cline{2-5}
    & 2D &$99.0$\% & $46.5$\%  & $98.75$\%  \\ \hline
    \multirow{2}{*}{Keller--Segel} & 1D & $100$\% &$100$\% & $100$\% \\ \cline{2-5}
    & 2D &$100$\% &$100$\% & $100$\%  \\ \hline
    \multirow{2}{*}{Biharmonic} & 1D & $99.43$\%& $100$\%& $99.43$\% \\ \cline{2-5}
    & 2D & $100$\% & $100$\% & $100$\% \\ \hline
    Nonlocal advection & 1D & $100$\% & $100$\% & $100$\% \\ \hline
    \end{tabular}
    \caption{Column 2: Proportion of simulations unpatterned at the final time $t=10^4$. Column 3: Proportion of simulations that were Turing unstable at the initial homogeneous equilibrium. Column 4: Proportion of the Turing-unstable simulations from Column 3 that decayed to a different homogeneous equilibrium.}\label{table_systematic_runs}
\end{table*}

Sufficiently changing the parameters of the models can give rise to other behaviours.
Rather than detail these observations, we encourage the reader to interactively explore the three local models with VisualPDE \citep{walker2023visualpde}\footnote{\url{https://visualpde.com/mathematical-biology/Turing-conditions-are-not-enough}}. For instance, by changing properties of the bistability, one can observe transitions between systems that form no patterns, favour localised solutions, or admit domain-filling patterns. Indeed, exploring the Keller-Segel equation interactively via VisualPDE is how we first observed this behaviour, with the other three models designed to mimic the basic ingredients of bistability and subcriticality.

\section{Discussion}\label{Discussion}
Across a range of models, parameter sets, and different initial conditions,   
we have robustly observed that possessing a Turing instability is not sufficient for systems to form spatial patterns that persist beyond transient timescales \ak{(the timescales observed in the examples in \cref{fig_1D_dispersion_sims} and \cref{fig_2D_sims} are plausibly too short to be compatible with many examples of biological patterning, though this would depend on the details of nondimensionalisation)}. Conversely, wave-pinning and other mechanisms can give rise to spatially structured stable states without a Turing-like bifurcation \citep{champneys2021bistability}. Therefore, while linear theory can have value in detecting self-organisation, it is perhaps not as generally valid as most of the textbook examples (e.g.~every reaction--diffusion system in the book  \citep{murray2003mathematical}) might indicate. 

\ak{We suspect that the almost ubiquitous association between Turing instabilities and pattern formation is largely because most research on patterning in reaction-transport systems, including our own \citep{krause2021modern}, focuses on small systems of at most three or four components with relatively mild nonlinearities. Systems such as \cref{RD_model} are still in this class of relatively simple systems, but the presence of bistability might be more indicative of large and complex reaction networks, which likely exhibit a high degree of multistability. Systematic analyses of such systems are relatively unexplored, and the results we have shown underscore the importance of studying them. Additionally, emphasis on supercritical bifurcations with stable small-amplitude patterns near the bifurcation point can fail to capture both systems exhibiting subcritical bifurcations as well as systems far away from the original Turing bifurcation point.}

\ak{Among modern tools for far-from-equilibrium analyses, we note that there exist several approaches for studying spike or pulse dynamics  \citep{wei2013mathematical, doelman2015explicit}. Such approaches have shown the importance of even small changes to the nonlinearity on the existence and stability of patterned states \citep{veerman2013pulses}. Contemporary numerical continuation techniques, such as in the \textsc{pde2path} software \citep{uecker2014pde2path} can be used to describe the loss of patterned states we explored here, as shown in Figure 11 of \cite{al2020bloom}. These approaches, however, typically focus on studying specific models and parameter sets, and do not lend themselves as easily to studying generic systems, especially those with more than two components. In contrast, linear stability theory has been employed to classify larger reaction-diffusion systems \citep{marcon2016high, scholes2019comprehensive, landge2020pattern}. Recent approaches such as Local Perturbation Analysis \citep{holmes2014efficient, holmes2015local} overcome some of these limitations, at the cost of only strictly applying in particular asymptotic regimes.  An important avenue would be the development of more powerful tools to understand complex and nonlinear systems in the context of pattern formation without relying on the limitations of looking only locally in the phase space or in the space of parameters/models. We view this as an exciting and important frontier for future theoretical work.}

\begin{acknowledgements}
 BJW is supported by the Royal Commission for the Exhibition of 1851.
\\
\item[\hskip\labelsep
\bfseries{Data Availability}]
There is no data presented in this paper. All code associated with the project can be found on GitHub \citep{Krause_Github}. Interactive versions of the local models can be found at the website \url{https://visualpde.com/mathematical-biology/Turing-conditions-are-not-enough}.
\end{acknowledgements}

\bibliographystyle{abbrvnat}
\bibliography{refs}

\end{document}